\documentclass[twocolumn,showpacs,preprintnumbers,amsmath,amssymb,aps,bibnotes,superscriptaddress]{revtex4}
\usepackage{graphicx}
\usepackage{color}

\renewcommand{\vec}[1]{\mathbf{#1}}

\newcommand{\vecgrk}[1]{\boldsymbol{#1}}
\newcommand{\eqnref}[1]{Eq.~\eqref{#1}}
\newcommand{\figref}[1]{Fig.~\ref{#1}}
\newcommand{\secref}[1]{Sec.~\ref{#1}}

\newcommand{\e}[1]{\text{e}^{#1}}
\newcommand{\cmplxi}{\text{i}}
\newcommand{\tr}{\operatorname{Tr}}
\newcommand{\punc}[1]{\,#1}
\newcommand{\neweqnline}{\nonumber\\}
\newcommand{\diffd}{\text{d}}

\begin{document}

\title{Itinerant ferromagnetism in an interacting Fermi gas with mass imbalance}

\author{C.W.~von Keyserlingk}
\affiliation{Rudolf Peierls Centre for Theoretical Physics, 1 Keble Road, Oxford,
OX1 3NP, United Kingdom}

\author{G.J.~Conduit}
\affiliation{Department of Condensed Matter Physics, Weizmann Institute of Science,
Rehovot, 76100, Israel}
\affiliation{Physics Department, Ben Gurion University, Beer Sheva, 84105, Israel}

\date{\today}

\begin{abstract}
We study the emergence of itinerant ferromagnetism in an ultra-cold atomic gas with a
variable mass ratio between the up and down spin species. Mass
imbalance breaks the SU(2) spin symmetry leading to a modified Stoner
criterion. We first elucidate the phase behavior in both the grand
canonical and canonical ensembles. Secondly, we apply the formalism to
a harmonic trap to demonstrate how a mass imbalance delivers unique
experimental signatures of ferromagnetism. These could help future
experiments to better identify the putative ferromagnetic
state. Furthermore, we highlight how a mass imbalance suppresses the
three-body loss processes that handicap the formation of a ferromagnetic
state. Finally, we study the time dependent formation of the
ferromagnetic phase following a quench in the interaction strength.
\end{abstract}

\pacs{03.75.Ss, 71.10.Ca, 67.85.-d}

\maketitle

\section{Introduction}

By exploiting the fine control of interactions through a magnetically
tuned Feshbach resonance~\cite{76s12}, ultracold atomic Fermi gases
have proven to be a rich arena in which to study many-body physics.
On one side of the Feshbach resonance the effective s-wave interaction
is attractive, which has allowed investigators to realize a BCS state
of Cooper pairs. If the interactions are tuned across the Feshbach
resonance, the fermions bind to form molecules that can subsequently
form a Bose-Einstein condensate~\cite{03sph08}. However, the fermions
experience repulsive interactions if the Feshbach resonance is
approached from the other side. A recent experiment
by the MIT group~\cite{Jo09} on the repulsive side of the resonance
has provided the first tentative evidence for the formation of an
itinerant ferromagnetic phase. There was however a significant atom
loss process that could drive the formation of alternative strongly
correlated states~\cite{Pekker10}, which are consistent with some of
the experimental results. It is therefore essential to develop a
different realization of the ferromagnetic state which suppresses
atom loss, and in addition delivers unique experimental
signatures to help resolve the outstanding questions over the original
experiment. If the experiment were
confirmed~\cite{Houbiers97,Duine05,LeBlanc09,Conduit10}, the
flexibility offered by ultracold atomic gases now presents
investigators with a unique opportunity to study aspects of
ferromagnetism that cannot be envisioned in the solid state, including
the consequences of population imbalance~\cite{Conduit08}, a conserved
net magnetization~\cite{Berdnikov09}, the damping of quantum
fluctuations by three-body loss~\cite{Conduit10ii}, single spin
flips~\cite{Zhai09}, spin drag~\cite{Duine10}, spin spiral
formation~\cite{Conduit-Altman}, reduced
dimensionality~\cite{Conduit-2D}, as well as ferromagnetic phenomena
in an optical lattice~\cite{FerroLattice}. In this paper we turn to consider
the consequences of the up and down-spin particles carrying different
masses, a scenario that cannot be realized in the solid state. Furthermore
the system offers investigators a lower atom loss rate combined with inimitable experimental
signatures.

The MIT experiment prepared an ultracold atomic gas with two different
atomic species to represent the pseudo up and down spin fermions. In
the experiment, and all theoretical studies of itinerant
ferromagnetism to date, the two species are different electronic
states of the same atom, and therefore carry the same mass. This
accurately reflects the situation in solid state ferromagnetism where
the up and down-spin electrons have the same effective mass. However,
the flexibility introduced by the new experimental realization of
ferromagnetism permits the pseudo up and down spins to be represented
by two species of atoms of different elements with unequal masses;
alternatively an optical lattice can change the effective mass of the
atoms. Though in the solid state interactions can change the effective mass of the
majority and minority spin species~\cite{Hirsch00}, distinct species
of fermions in an ultracold atomic gas provide a cleaner and more
controlled realization of mass imbalance. Furthermore, introducing a mass ratio
breaks the SU(2) symmetry of the conventional ferromagnet, and as a
result the magnetization has anisotropic susceptibility and so offers a
novel control parameter over magnetic ordering and unique
experimental signatures of the ferromagnetic state. Moreover,
introducing a mass imbalance should suppress the three-body losses
that hinder ultracold atom experiments~\cite{Petrov03}. Further
motivation to study a generalized mass ratio stems from novel physics
discovered on the attractive side of the Feshbach resonance. It has
been established that imbalanced Fermi surfaces in a superfluid can
drive the formation of the textured Fulde-Ferrel-Larkin-Ovchinnikov
phase~\cite{FFLOPapers}, and here we explore the opportunity that new
phenomena could arise in a mass imbalanced itinerant ferromagnet.

In this paper in \secref{sec:Formalism} we first develop the formalism
required to study a mass imbalanced Fermi gas with repulsive
interactions. Subsequently in \secref{sec:Phase-Diagram} we derive the
general phase diagram for a uniform gas with both a generalized mass
ratio and also population imbalance. To cement the connection to the
recent possible experimental observation of ferromagnetism in an
ultracold atom gas, we consider the consequences of a trapped geometry
and study the quantities observable by experiments in
\secref{sec:TrappedGeometry}. In the current experimental realization
of ferromagnetism there were significant three-body losses so the
ferromagnetic phase was formed following a quench in the interaction
strength out of equilibrium. Therefore in \secref{sec:TextureandPerp} we
conclude our investigation by studying the dynamical formation of the
ferromagnetic phase. Finally, we summarize our discussion of itinerant
ferromagnetism in
\secref{sec:Conclusions}.

\section{Formalism}\label{sec:Formalism}

To study itinerant ferromagnetism in the presence of a mass imbalance
we use the functional integral formalism developed in
Ref.~\cite{Conduit08}. The phase diagram predicted there has recently
been verified by \emph{ab initio} quantum Monte Carlo
studies~\cite{Conduit09,Pilati10} and is also in accord with recent
experimental findings~\cite{Jo09,LeBlanc09,Conduit10}. Therefore, the
approach described in Ref.~\cite{Conduit08} provides a solid platform
from which to investigate itinerant ferromagnetism with a mass imbalance.

The formalism centers around calculating the quantum partition
function expressed as a coherent state field integral
\begin{align}
\mathcal{Z}\!=\!\!\int\!\!\mathcal{D}\psi\text{exp}\Bigg[\!-\!\!\int\!\!\!\!\!\sum_{\sigma=\left\{\uparrow,\downarrow\right\}}\!\!\!\!
\overline{\psi}_{\sigma}(\!\hat{\partial}_{\tau}\!+\!\hat{\xi}_{\sigma}\!)\psi_{\sigma}\!-\!g\!\int \!\overline{\psi}_{\uparrow}\overline{\psi}_{\downarrow}\psi_{\downarrow}\psi_{\uparrow}\Bigg]\punc{.}
\label{eq:partition}
\end{align}
Here the field $\psi$ describes a two component Fermi gas
with a repulsive s-wave contact interaction $g\delta^{3}(\vec{r})$
acting between the two species. We use the notation
$\int\equiv\int_{0}^{\beta}\diffd\tau\int\diffd\vec{r}$ with inverse
temperature $\beta=1/k_{\text{B}}T$, and dispersion
$\hat{\xi}_{\sigma}=\hat{p}^{2}/2m_{\sigma}-\mu_{\sigma}$. It will
later be convenient to rewrite the particle masses as
$m_{\sigma}=m(1+\sigma r)$, and the species chemical potentials as
$\mu_{\sigma}=\mu+\sigma\Delta\mu$. Here
$\sigma\in\{\uparrow,\downarrow\}$ is a label to distinguish between
the two species of atoms and does not represent a physical spin.

We now decouple the quartic interaction term, which will allow us to
integrate out the fermionic degrees of freedom. Hertz did this by
introducing a scalar Hubbard-Stratonovich decoupling of the two-body
interaction term into the magnetization channel~\cite{Hertz}. By
expanding in the magnetization he was able to develop an effective
Landau theory. However, recent studies have shown that this approach
fails to recover the correct Hartree-Fock equations, and capture the
behavior of the soft transverse degrees of
freedom~\cite{Conduit08}. Moreover, as mass imbalance breaks the SU(2)
symmetry of the system, magnetization formed along the quantization
axis is distinctly different from perpendicular magnetization. We
therefore adapt the formalism developed in Ref.~\cite{Conduit08}, and
decouple the quartic interaction term in the full vector magnetization
$\vecgrk{\phi}$ as well as the density channel $\rho$. This yields an
action that is quadratic in the fermion degrees of freedom, and after
integrating them out we recover the quantum partition function
$\mathcal{Z}=\int\mathcal{D}\vecgrk{\phi}\mathcal{D}\rho\exp(-S)$ with
an action
\begin{align}
 S\!=\!\int\!g(\vecgrk{\phi}^{2}\!-\!\rho^{2})\!
 -\!\tr\ln\left[(\hat{\partial}_{\tau}\!+\!\hat{\xi}_{\alpha}\!+\!g\rho)\mathbf{I}-g\vecgrk{\phi}\cdot\vecgrk{\sigma}\right]\punc{.}
 \label{eq:phiaction}
\end{align}

We now focus on the saddle point fields (or ``mean-fields'') of the
action, that is $\vecgrk{\phi}$ and $\rho$ satisfying $\delta
S/\delta\vecgrk{\phi}=0$ and $\delta S/\delta\rho=0$. We show in \secref{sec:PerpPol} that
fluctuations are gapped so we
neglect the fluctuation corrections and assume that the saddle point
fields make the dominant contribution to the partition function. We then diagonalize the inverse Green
function inside the trace to the new basis set $\chi\in\{+,-\}$, and
perform the summation over Matsubara frequencies. Finally we use
$\Phi=-k_{\text{B}}T\ln(\mathcal{Z})$ to yield the thermodynamic grand
potential
\begin{align}
 \Phi\!=\!gV(\vecgrk{\phi}^{2}\!-\!\rho^{2})\!-\!k_{\text{B}}T\!\!\!\!\!\sum_{\chi\in\{+,-\}}\int\!\diffd\epsilon\,\nu(\epsilon)\ln\left(1\!+\!\e{-\beta\zeta_{\chi}}\right)\punc{,}
 \label{eq:finalaction}
\end{align}
where $V$ is the total volume,
$\nu(\epsilon)=m^{3/2}\sqrt{\epsilon}/\pi^{2}\hbar^{3}\sqrt{2}$ is the
density of states, and effective dispersion
\begin{align}
 \zeta_{\pm}\!=\!\frac{\epsilon}{1-r^{2}}\!-\!\mu\!
 +\!g\rho\pm\sqrt{(g\phi_{\perp})^{2}\!+\!\left(\frac{\epsilon r}{1-r^{2}}\!+\!\Delta\mu\!+\! g\phi_{\text{z}}\right)^{2}}\punc{.}
\end{align}

Varying the grand potential $\Phi$ with respect to $\vecgrk{\phi}$ and $\rho$ yields
the saddle point equations for the homogeneous mean-fields. We have
checked numerically in \secref{sec:Phase-Diagram} and analytically in
\secref{sec:PerpPol} that any gas with a mass and/or chemical
potential imbalance breaks the SU(2) symmetry and does not develop
perpendicular magnetization, so $\phi_{\perp}=0$. This result
considerably simplifies the mean-field equations. As the particle
densities are related to the saddle point fields through
$n_{\sigma}=\rho+\sigma\phi_{\text{z}}$, at zero temperature we cast
the mean-field equations as
\begin{align}
 n_{\sigma}=\frac{\sqrt{2}}{3\pi^{2}\hbar^{3}}m_{\sigma}^{3/2}\left(\mu_{\sigma}-gn_{-\sigma}\right)^{3/2}\punc{,}
 \label{eq:master}
\end{align}
which shows that the Fermi energy of the $\sigma$ species is
$\epsilon_{\text{F}\sigma}=\mu_{\sigma}-gn_{-\sigma}$. This equation
forms the backbone of our subsequent analysis. We first use it to
derive a generalized Stoner criterion for the ferromagnetic
instability in a mass imbalanced gas. Such a transition is
characterized by the appearance of three nearby solutions; the now unstable
original state, and two with small relative positive and negative
polarization. Demanding the existence of these solutions to the
self-consistent equations \eqref{eq:master} yields a modified Stoner criterion
$g\sqrt{\nu_{\uparrow}\nu_{\downarrow}}=1$, where the $\nu_{\sigma}$
is a density of states at the $\sigma$ species Fermi surface. This
reduces to the familiar criterion $g\nu=1$ in the mass and population
balanced limit~\cite{Stoner}.

In \secref{sec:TexturedPhase} we show that the saddle point fields are
always uniform, even for general chemical potential and mass
imbalance. Therefore, unlike the superfluid regime where a Fermi
surface imbalance can drive the formation of the textured
Fulde-Ferrel-Larkin-Ovchinnikov state
\cite{FFLOPapers}, here for a perfectly spherical Fermi surface and
the absence of nesting only uniform ferromagnetic states will be
formed. We note however that fluctuation corrections drive textured
phase formation in the equal mass case~\cite{Conduit09}, and have
the potential to do the same with mass imbalance.

\section{Phase Diagram\label{sec:Phase-Diagram}}

Now that we have prepared the formalism we are ideally placed to study
the phase diagram of the mass imbalanced Fermi gas with repulsive
interactions. This will allow us to build up an intuition for the
consequences of mass imbalance before we turn to study the gas in a
harmonic trap in \secref{sec:TrappedGeometry}. Firstly, in
\secref{sec:GrandCanonical}, we examine the grand canonical ensemble
with a gas connected to an infinite particle reservoir, and secondly in
\secref{sec:Canonical} we study the gas with a constant number of
particles in the canonical ensemble. To cement the connection to
experiment, from now on we express the interaction strength as
$k_{\text{F}}a$, the product of the Fermi wave vector $k_{\text{F}}$
of a non-interacting gas with the same net number density, and the
scattering length $a=g/2\pi\hbar^{2}(1/m_{\uparrow}+1/m_{\downarrow})+\mathcal{O}(g^{2})$~\cite{Meystre01}.
Consistent with the Hartree-Fock
scheme we employed to calculate the grand thermodynamic potential \eqnref{eq:finalaction}
we have taken the lowest order term in the scattering
length and neglected higher order corrections in $g$~\cite{Duine05,Conduit08,Zhou11}.
We concentrate on the phase behavior at $T=0$.

\subsection{Grand canonical ensemble}\label{sec:GrandCanonical}

\begin{figure}
 \includegraphics[width=0.9\linewidth]{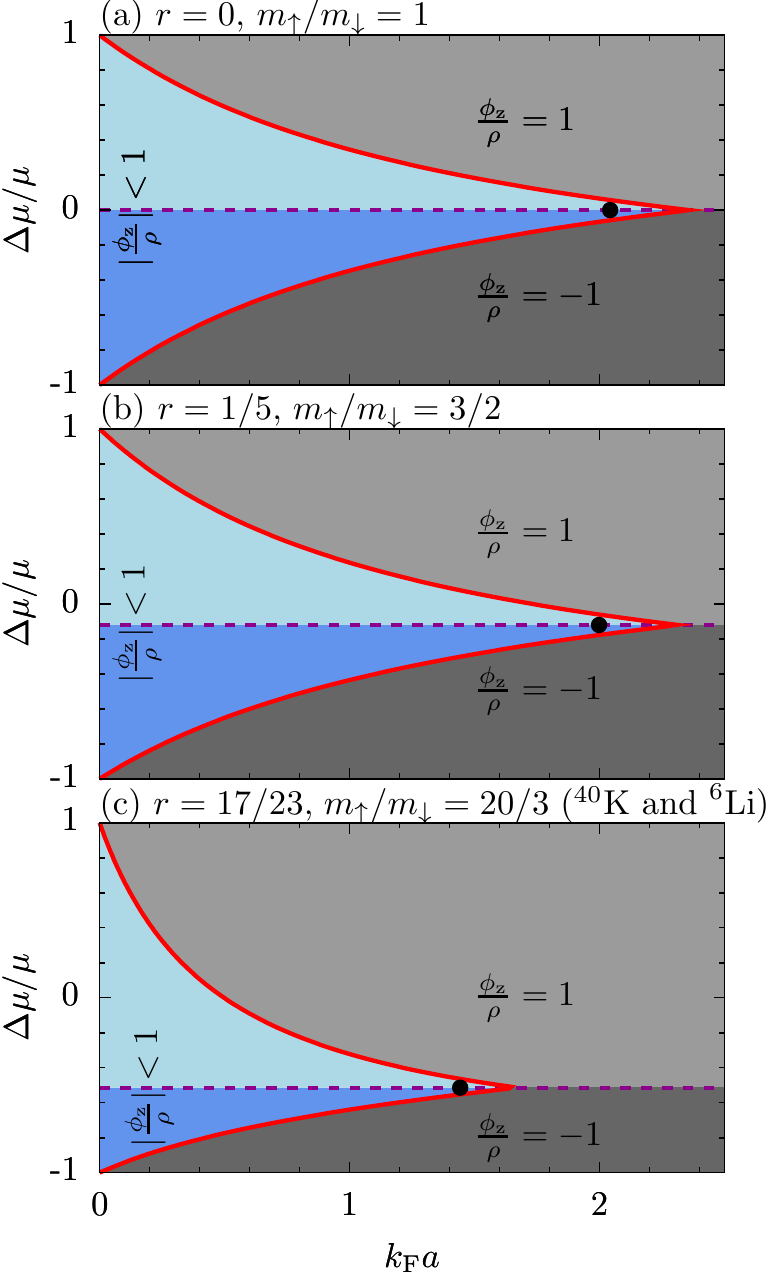}
 \caption{(Color online) The phase behavior in the grand canonical ensemble with
 chemical potential imbalance $\Delta\mu/\mu$ and interaction strength
 $k_{\text{F}}a$ for three different values of mass imbalance. The solid red
 lines denote the onset of full polarization, into the phases shaded
 light and dark gray for up ($\phi_{\text{z}}/\rho=1$) and down
 ($\phi_{\text{z}}/\rho=-1$) spin respectively. The dashed purple
 boundary separates systems that become polarized in the up (above
 dashed line, light blue shading) or down-spin (below dashed line,
 dark blue shading) directions in the strongly interacting limit.
 Along the dashed purple line itself the
 polarization remains constant until reaching the black dot, past
 which polarization in either direction is equally favorable.}
 \label{Flo:2dzphase}
\end{figure}

In the grand canonical ensemble the gas can exchange particles with an
ideal reservoir. We can control the average number density of
atoms by manipulating the chemical potentials $\mu_{\sigma}$ of
the reservoir.

To develop a physical understanding and connect with previous
work~\cite{LeBlanc09,Conduit08,Conduit10} we first describe the phase
diagram for a gas without a mass imbalance shown in
\figref{Flo:2dzphase}(a). When the gas is noninteracting it is
paramagnetic for all chemical potential imbalances
$\Delta\mu=\mu_{\uparrow}-\mu_{\downarrow}$. Above the line
$\Delta\mu=0$, increasing the interaction strength drives polarization
in the up-spin direction. Conversely, below $\Delta\mu=0$ the gas
becomes polarized in the down-spin direction. As the interaction
strength is increased across the boundary marked on
\figref{Flo:2dzphase}(a) the gas enters the fully polarized state. For
more positive $\Delta\mu$, the gas becomes fully polarized in the
$\uparrow$ direction at lower interaction strength. We can understand
the form of the boundary line at $\Delta\mu>0$ by examining the
species Fermi energies
$\epsilon_{\text{F}\sigma}=\mu+\sigma\Delta\mu-gn_{-\sigma}$. At $g=0$
the Fermi energies are simply the chemical potentials, so
$n_{\uparrow}>n_{\downarrow}$. The result is that
$\epsilon_{\text{F}\downarrow}$ is smaller than
$\epsilon_{\text{F}\uparrow}$ at zero interaction, and decreases more
rapidly with increasing interaction strength. Thus the gas becomes
fully polarized more quickly in the $\uparrow$ direction as we
increase the chemical potential bias. An analogous
situation occurs in the bottom half of
\figref{Flo:2dzphase}(a), where $\uparrow$ and $\downarrow$ swap
roles in the argument above. The line separating these two regimes is $\Delta\mu=0$. As the
interaction strength is increased along this line the magnetization
remains pinned at zero until $k_{\text{F}}a\approx2.04$,
at which point a ferromagnetic instability develops and
the gas can become polarized in any direction. The instability to full
polarization at $k_{\text{F}}a=3\pi/4$ coincides with the cusped
junction of the fully polarized region boundary.

When we introduce a mass imbalance, for any chemical potential and
interaction strength, the saddle point solutions have
$\phi_{\perp}=0$. The phase behavior shown in
\figref{Flo:2dzphase}(b and c) is then obtained using the zero
temperature mean-field equations \eqref{eq:master}. At zero interaction
strength when there is no chemical potential imbalance, the number
density of the species is
$n_{\sigma}=(\sqrt{2}/3\pi^{2}\hbar^{3})m_{\sigma}^{3/2}\mu^{3/2}$,
and so an increase in the mass imbalance
alone will bias the system towards the heavy spin species. Then, as we
increase the interaction strength, the Fermi energy
$\epsilon_{\text{F}\downarrow}=\mu-gn_{\uparrow}$ of the minority
lighter species will fall more rapidly than that of the heavy species
$\epsilon_{\text{F}\uparrow}=\mu-gn_{\downarrow}$, and so the gas
becomes fully polarized towards the heavier species. With the heavier
species becoming favored, the border in \figref{Flo:2dzphase}(b and c)
between the dominant heavy and light spin polarization shifts downwards towards
the lighter spin particles. In order to have neither species dominant
at full polarization, we need to introduce a chemical potential
imbalance that favors the lighter species, given by the implicit
equation
\begin{align}
 \frac{\Delta\mu}{\mu}\!=\! \frac{[(1\!-\!r)(\mu\!-\!\Delta\mu)]^{3/2}
 \!-\![(1\!+\!r)(\mu\!+\!\Delta\mu)]^{3/2}}{[(1\!-\!r)(\mu\!-\!\Delta\mu)]^{3/2}
 \!+\![(1\!+\!r)(\mu\!+\!\Delta\mu)]^{3/2}}\punc{.}
 \label{eq:implicit}
\end{align}

If the chemical potentials are tuned according to
\eqnref{eq:implicit}, then at low interaction strength the magnetization is
pinned to $-\Delta\mu/\mu$. Increasing the interaction strength past
a critical value (denoted by the black dot in \figref{Flo:2dzphase})
induces a second order phase transition in the magnetization; the
minimum in the grand potential at $-\Delta\mu/\mu$ bifurcates into two
equally favorable minima which move continuously to full up and down
polarization as we further increase the interaction strength.
Full polarization emerges at an interaction strength
\begin{align}
k_{\text{F}}a = \frac{3\pi\hbar k_{\text{F}}}{2\sqrt{2}\left(1/m_{\uparrow}+1/m_{\downarrow}\right)}\max_{\sigma}\left(\frac{\mu_{-\sigma}}{(m_{\sigma}\mu_{\sigma})^{3/2}}\right)\punc{.}
 \label{eq:gfp}
\end{align}

Having understood the behavior of the gas in the grand canonical
ensemble we now disconnect the particle reservoirs and study a gas in
the canonical ensemble.

\subsection{Canonical ensemble}\label{sec:Canonical}

\begin{figure}
 \includegraphics[width=0.9\linewidth]{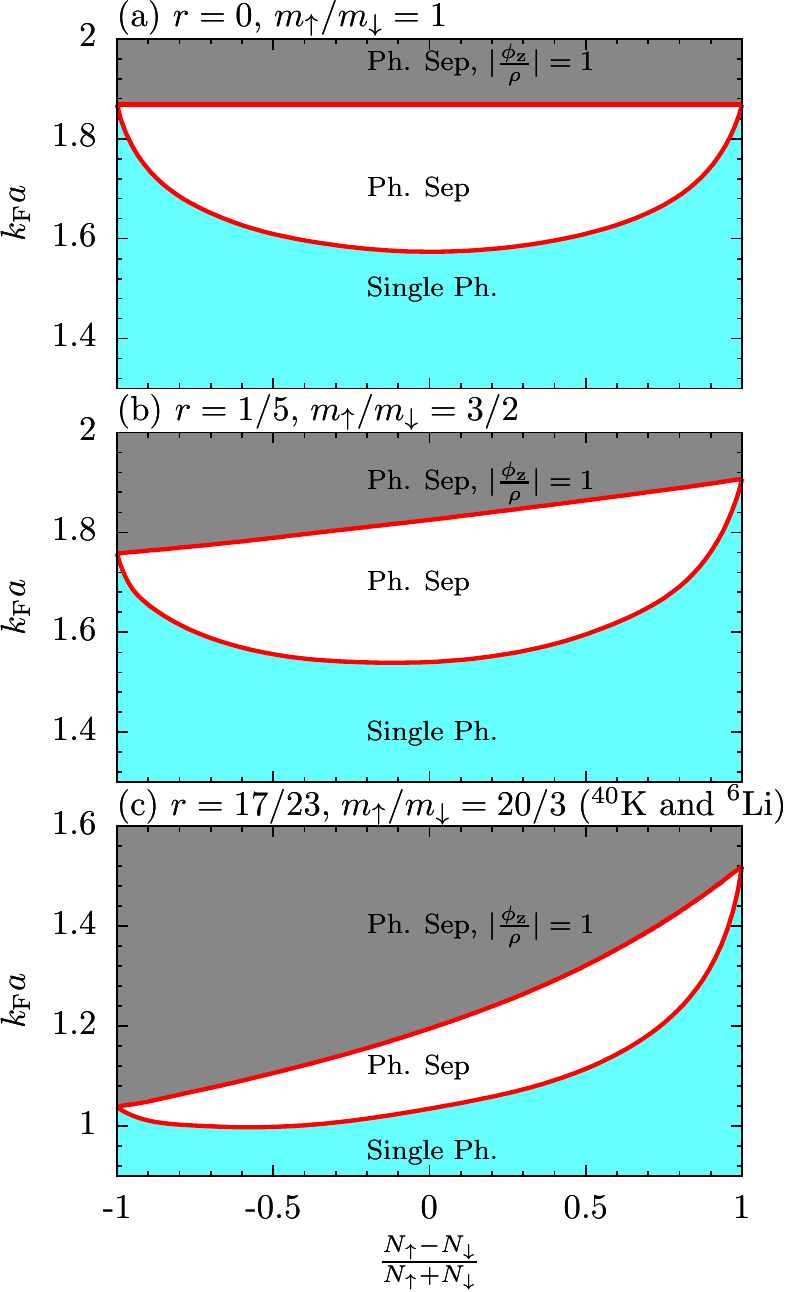}
 \caption{(Color online) The phase behavior in the canonical
 ensemble with interaction strength $k_{\text{F}}a$ and net
 population imbalance
 $(N_{\uparrow}-N_{\downarrow})/(N_{\uparrow}+N_{\downarrow})$ for
 three different values of mass imbalance $r$. The
 light blue regions (Single Ph.) correspond to a paramagnetic
 gas, white (Ph. Sep)
 to a two-phase coexistence between two partially polarized gases,
 and dark gray (Ph. Sep, $|\frac{\phi_{\text{z}}}{\rho}|=1$)
 to a two-phase coexistence between two fully polarized up and down
 spin gases. } \label{Flo:Canonical}
\end{figure}

We now investigate a gas confined so that the total number of both
species is held fixed. In a cold atom gas the number of up and down
particles are separately conserved, and so at the ferromagnetic
transition the gas splits into up and down polarized domains. On
increasing the interaction strength the gas could potentially phase
separate into majority up and majority down-spin domains to reduce
contact between the two species. The formation of this
ferromagnetic state is governed by the competition between the
resulting fall in interaction energy and a kinetic energy penalty due to
the increased density of each separate species.

A box of gas with fixed total particle numbers is described by
minimizing the total free energy. The free energy density of a single
domain with particle densities $n_{\uparrow}$ and $n_{\downarrow}$ is
obtained from the grand potential \eqnref{eq:finalaction} by
substituting our mean-field solutions \eqnref{eq:master} into the
definition
$F=\Phi+\mu_{\uparrow}n_{\uparrow}V+\mu_{\downarrow}n_{\downarrow}V$,
which yields
\begin{align}
 \frac{F}{V}=\frac{3}{5}\left(\frac{3\pi^{2}\hbar^{3}}{\sqrt{2}}\right)^{2/3}\left[\frac{n_{\uparrow}^{5/3}}{m_{\uparrow}}+\frac{n_{\downarrow}^{5/3}}{m_{\downarrow}}\right]
 +gn_{\uparrow}n_{\downarrow}\punc{.}
\end{align}
If there is no phase separation, the total free energy of the gas is
just that of a single domain. This state competes with a
phase separated gas containing two domains, labeled A and B,
with a ratio of volumes $\gamma$, and particle densities
$n_{\text{A}\sigma}$ and $n_{\text{B}\sigma}$. The total energy of the
state is then $F=\gamma F_{\text{A}}+(1-\gamma)F_{\text{B}}$, and the
total particle numbers are $N_{\sigma}=V\left(\gamma
n_{\text{A}\sigma}+\left(1-\gamma\right)n_{\text{B}\sigma}\right)$.
We then minimize the total free energy while fixing the $N_{\sigma}$,
to determine whether the system phase separates, and if so the
properties of the individual domains. The resulting phase diagrams are
shown in \figref{Flo:Canonical}.

To explore the phase behavior in the canonical ensemble, and to
establish a connection to the literature \cite{Conduit08}, we first
focus on the mass balanced case shown in
Fig.\ref{Flo:Canonical}(a). At weak interactions the gas starts in the
paramagnetic state. On ramping the interaction strength through the
Stoner criterion at $k_{\text{F}}a=\pi/2$, a system with zero net
population imbalance phase separates into two weakly but oppositely
polarized domains. The critical interaction strength for phase
separation is higher in the presence of a population imbalance due to the larger
kinetic energy barrier that must be overcome for further polarization
to form. A fully polarized phase forms at
$k_{\text{F}}a=3\pi/2^{7/3}$, which is in accordance with
Ref.~\cite{Conduit08,LeBlanc09}.

On introducing a mass imbalance the phase diagrams tilt so that when
the population imbalance is towards the lighter species, the onset of
full phase separation takes place at smaller $k_{\text{F}}a$. To
understand this, we first derive an expression for the interaction
strength at which the system becomes fully polarized. At this point,
there is an A phase composed entirely of the heavier $\uparrow$
particles and a B phase of the lighter $\downarrow$ particles. Just
before the transition to full polarization, there will still be a
$\downarrow$ particle in the A phase. This has interaction energy
$g_{\text{fp}}n_{\uparrow}$. If that atom transits into phase B it sits on top of
the Fermi surface, with an energy penalty
$D n_{\downarrow}^{2/3}/m_{\downarrow}$, where
$D=3^{5/3}\pi^{4/3}\hbar^{2}/2^{1/3}5$. At the transition to full
polarization the particle makes this passage without hindrance, and
therefore these two energies are equal $g_{\text{fp}}n_{\uparrow}=Dn_{\downarrow}^{2/3}/m_{\downarrow}$.
This implies that the interaction strength at the phase transition is
$g_{\text{fp}}=D n_{\downarrow}^{2/3}/m_{\downarrow}n_{\uparrow}=D n_{\uparrow}^{2/3}/m_{\uparrow}n_{\downarrow}$;
to deduce the second equality we have repeated the argument for down
spin particles. The second equality implies that
$n_{\downarrow}^{5/3}/m_{\downarrow}=n_{\uparrow}^{5/3}/m_{\uparrow}$,
which confirms that the two regions have equal pressure.

We now use the above expressions to explain the tilt in the fully
polarized phase boundary \figref{Flo:Canonical}(b) and (c). To go from
population balance to a system imbalanced towards the light particles,
we imagine converting a region of heavy particles into light. However,
for a given density, the lighter particles exert a greater degeneracy
pressure $P\propto n^{5/3}/m$ than heavier particles and so the
internal pressure within the system must increase. This compresses
both the light and heavy particle domains, and since
$n_{\downarrow}^{5/3}/m_{\downarrow}=n_{\uparrow}^{5/3}/m_{\uparrow}$
the overall density of both phases must increase by the same
ratio. Therefore, upon biasing the population towards the lighter species
the critical interaction strength $g_{\text{fp}}\propto
n_{\downarrow}^{2/3}/n_{\uparrow}$ must decrease, thus tilting the
phase boundary.

\section{Trapped geometry \& experimental observation}\label{sec:TrappedGeometry}

\begin{figure*}
 \includegraphics[width=0.9\linewidth]{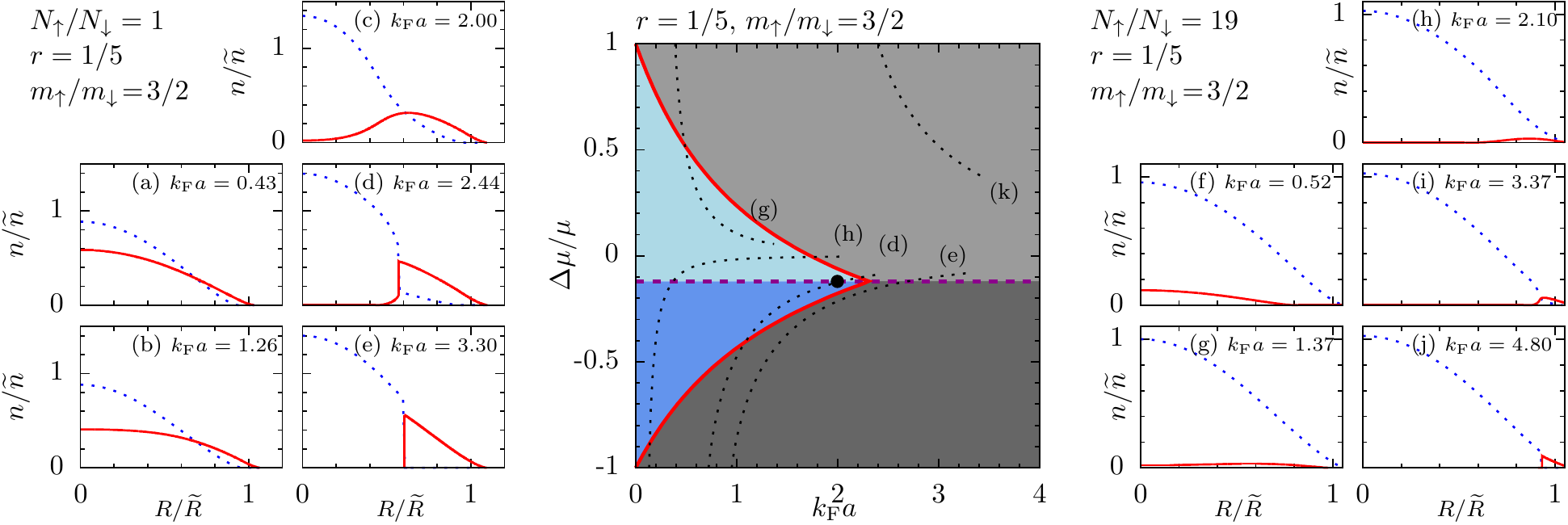}
 \caption{(Color online) Trap-profiles for a gas at various interaction strengths
 with a mass imbalance $m_{\uparrow}/m_{\downarrow}=3/2$. Figures (a to e)
 for a cloud with $N_{\uparrow}/N_{\downarrow}=1$ show the
 density of the heavier particles (dotted) and the density of lighter
 particles (solid) against radius. The axes are normalized by
 $\widetilde{n}$, the density of the heavier particles at the center
 of the trap when $k_{\text{F}}a=0$, and $\widetilde{R}$, the cloud
 size when $k_{\text{F}}a=0$. Figures (f to j) show the density profiles
 for a cloud with $(N_{\uparrow}-N_{\downarrow})/(N_{\uparrow}+N_{\downarrow})=0.9$.
 At the center a grand canonical phase diagram reproduced
 from \figref{Flo:2dzphase} shows dotted trajectories in parameter
 space corresponding to some of the trap-profiles and (k) is referenced in the text.} \label{Flo:traj}
\end{figure*}

Having understood the behavior in the canonical and grand canonical
ensembles in a uniform background potential, we are well positioned to
study the experimental realization of the gas in a harmonic trap
potential $V(R)=\omega R^{2}$. We employ the local-density
approximation and so assume that the properties of the gas at radius
$R$ are determined by substituting a local chemical potential
$\mu_{\sigma}(R)=\mu_{\sigma}-V(R)$ into our mean-field relations
\eqnref{eq:master}. Moving outwards from the center of the trap, the
system parameters trace trajectories on the grand canonical phase
diagrams \figref{Flo:2dzphase}. An immediate corollary is that the gas
becomes polarized only along the quantization axis, and so the gas
separates into domains of light and heavy particles. The chemical
potentials $\mu_{\sigma}$ at the center of the trap are chosen to
ensure that the cloud contains a fixed total number of atoms, and then
the properties of the gas are entirely determined by the interaction
strength and particle masses. In what follows, we will be interested
in four main trap observables: the number density profiles of the two
species, the total trap-size, and loss rate, which can all be measured
by imaging the spatial distribution of the atoms \textit{in situ}, and
the total kinetic energy, measured by tracking the profile of the
atoms following a ballistic expansion. Following~\cite{Petrov03}, we
model the loss rate density according to $\Upsilon n_{\uparrow}n_{\downarrow}
a^{6}$, where $\Upsilon$ contains the residual mass and number density
terms. $\Upsilon$ has a non-monotonic mass ratio
dependence~\cite{Petrov03}, leading to a dramatic suppression of loss
rate for moderately large mass imbalances (for example a
gas containing $^{40}$K and $^{6}$Li has
$m_{\uparrow}/m_{\downarrow}=20/3$, which we show reduces
loss by at least a factor of 20).

We are interested in how mass imbalance affects these four
experimental observables, but to contrast our results with earlier
work, we first review the mass balanced case. After that, we introduce
mass imbalance and expose unique signatures of the ferromagnetic
phase.

\subsection{Mass balanced gas}

We start by examining a trap with a two component Fermi gas with mass
balance, but variable population imbalance. To understand the
corresponding trends in \figref{Flo:array} (comparing solid red curves
in the same row), it is first useful to note how the species are
redistributed in the trap as we increase the repulsive interaction strength.

\emph{Density profiles}: At zero interaction, both species have identical
smooth distributions in the trap. Upon increasing $k_{\text{F}}a$, the
species spread themselves more thinly across the trap to reduce
repulsion. As the interaction strength passes a critical threshold,
magnetic domains are formed in the center of the trap via a
spontaneous symmetry breaking~\cite{Babadi09}. However, because the
density (and hence effective interaction strength) decreases at larger
radii, the gas
remains paramagnetic there. The width of this outer
paramagnetic region falls as $\sim 1/(k_{\text{F}}a)^{2}$ in the large
$k_{\text{F}}a$ limit. On introducing a population imbalance, some of
the minority spin particles are driven to larger radii with increasing
$k_{\text{F}}a$. This is because the minority species feels an
interaction energy proportional to the density of the majority
species, which overcomes the trapping potential. For large enough
$k_{\text{F}}a$, domains form in the central regions of the
trap. These become fully polarized as the interaction strength
continues to increase. In this fully polarized limit there is no
overlap between the species so the interaction energy disappears and
both species can be found in domains anywhere across the trap.

\emph{Cloud size}: In the non-interacting limit, a cloud with a population imbalance
contains more of the majority spin species so has a greater initial
radius. Initially, cloud size increases linearly at small
$k_{\text{F}}a$ as the atoms repel and spread themselves more sparsely
through the trap. After the atoms enter ferromagnetic domains, firstly
at the center and later across the entire trap, the cloud size
asymptotes towards its large $k_{\text{F}}a$ limit. At strong
interactions the fully polarized domains contain effectively a
non-interacting gas and so the cloud size is the same as for the
population balanced case.

\emph{Kinetic energy}: At zero interaction strength the total kinetic
energy of each species is $E_{\text{K}\sigma}=2^{-13/6}3^{4/3}\omega^{1/2}\hbar N_{\sigma}^{4/3}/m^{1/2}$.
A population imbalance deposits particles on top of the majority spin
Fermi surface increasing its kinetic energy, whilst the minority
spin species kinetic energy falls. The mass-balanced curves in
\figref{Flo:array} show an initial decrease in kinetic energy against
$k_{\text{F}}a$ as the atoms repel and spread themselves more thinly
across the trap. However, the onset of ferromagnetism drives up the
kinetic energy because identical fermions are confined at higher
densities within polarized domains. At strong interactions the gas
separates into independent fully polarized domains so the
kinetic energy of the cloud plateaus out.

\emph{Loss rate}: The loss rate $\Upsilon n_{\uparrow}n_{\downarrow} a^{6}$
initially rises with
interaction strength since it is proportional to
$\left(k_{\text{F}}a\right)^{6}$. At the ferromagnetic transition the
two species are confined to separate domains suppressing the factor
$n_{\uparrow}n_{\downarrow}$ and the loss rate falls. Introducing an
initial population imbalance also reduces the factor
$n_{\uparrow}n_{\downarrow}$, resulting in a lower loss rate at all
interaction strengths.

\subsection{Mass imbalanced gas}

\begin{figure*}
 \includegraphics[width=0.9\linewidth]{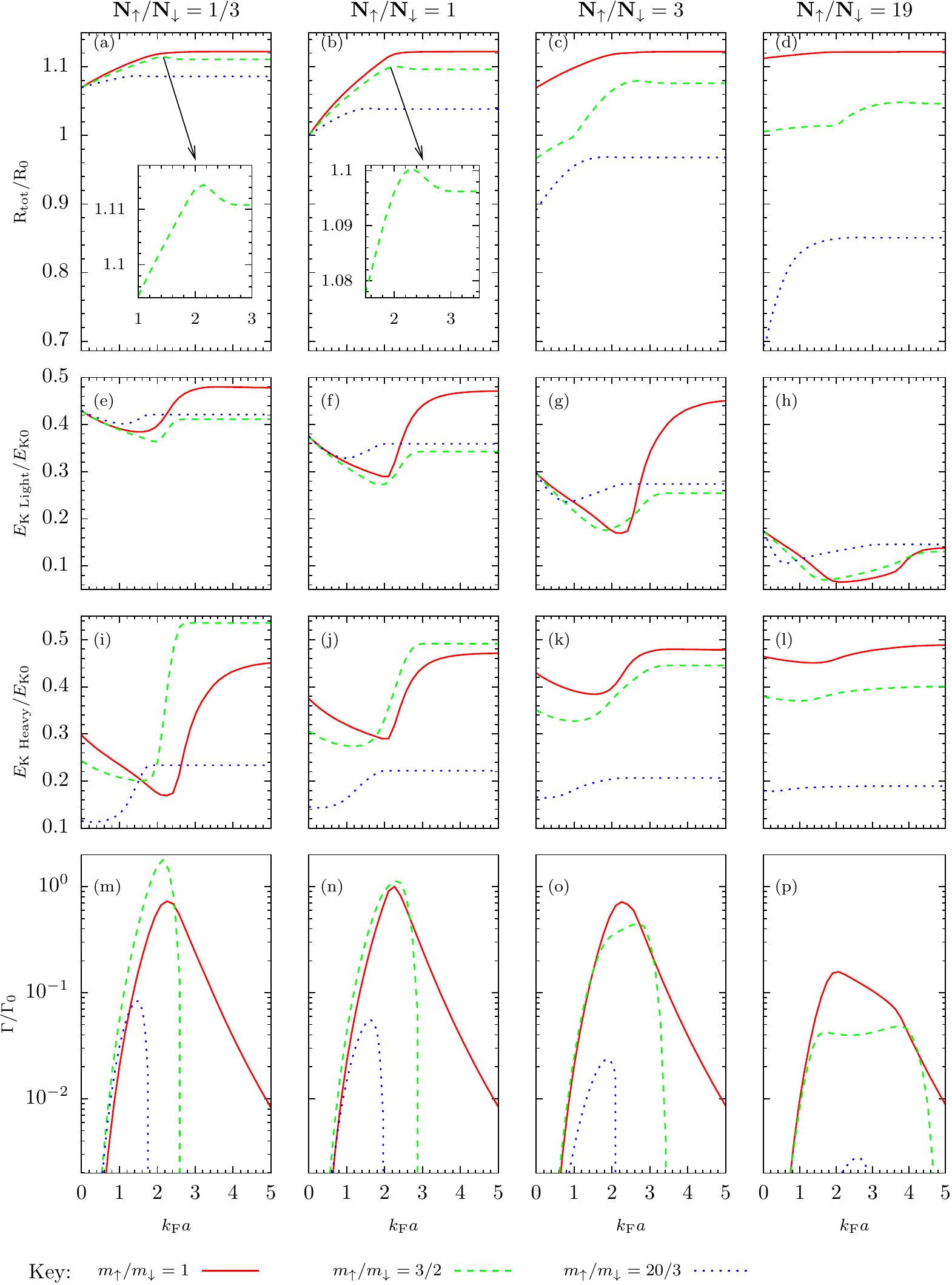}
 \caption{(Color online) A table of graphs with rows showing the cloud
 size ($R_{\text{tot}}$), kinetic energy ($E_{\text{K}}$) per particle
 of light and heavy species, and loss rate ($\Gamma$) as a function of
 interaction strength. The columns refer to population imbalances of
 $N_{\uparrow}/N_{\downarrow}=\{1/3,1,3,19\}$. Each graph shows three
 different mass imbalances $m_{\uparrow}/m_{\downarrow}=1$ by the solid red
 line, $m_{\uparrow}/m_{\downarrow}=3/2$ by the green dashed line,
 and the case of $^{40}$K and $^6$Li,
 $m_{\uparrow}/m_{\downarrow}=20/3$, is the blue dotted line. The axes
 are normalized by the cloud size ($R_{0}$), kinetic energy of a
 particle at R=0 ($E_{\text{K} 0}$), and peak loss rate ($\Gamma_{0}$)
 for a mass and population balanced non-interacting cloud with the
 same total particle number.} \label{Flo:array}
\end{figure*}

Having understood the behavior of a trapped gas with mass balance, we
now have a firm platform from which to study a gas with mass
imbalance. The imbalance is introduced by replacing the $\uparrow$
species with a more massive particle, while keeping the mass of the
$\downarrow$ species the same. In \figref{Flo:array} and below we
catalog and analyze the resulting changes that could offer
experimentalists both a handle to reduce losses, and new unique
signatures of ferromagnetic ordering.

\subsubsection*{Density profiles}

In \figref{Flo:traj} we examine the density profiles of the trapped
atomic gas with a mass imbalance of $m_{\uparrow}/m_{\downarrow}=3/2$.
In \figref{Flo:traj}(a) at small interaction strength each species is
supported within the cloud mostly by its own internal Fermi degeneracy
pressure
$\hbar^{2}(6\pi^{2})^{2/3}n_{\sigma}^{5/3}/5m_{\sigma}$. As the
degeneracy pressure is lower for the heavy species they are
denser at the trap center than the light species. Increasing the
interaction strength through \figref{Flo:traj}(b and c) expels the
lighter particles to larger radii, while the heavy particles become
more concentrated at the center. Raising the interaction strength
still further in \figref{Flo:traj}(d) leads to full polarization of
the heavy particles in the center of the trap, full light particle
polarization at the edge of the trap, and an intermediate density
discontinuity. At this point the trajectory (d) in \figref{Flo:traj}
passes through the point denoting the ferromagnetic instability.
By \figref{Flo:traj}(e) the entire gas has become
fully polarized at $k_{\text{F}}a\approx3.3$, whereas in the mass
balanced case the gas never became fully polarized. To understand why the heavy particles
congregate at the trap center, imagine instead that the heavy
particles come to dominate the outer regions of the trap. This would
require the chemical potential of the heavier species to be larger
than that of the lighter species. However, looking at the phase
diagram \figref{Flo:traj}, any trajectory (k) with positive chemical potential
imbalance curves upwards which implies that the heavier particles
dominate the whole of the trap in the full polarized limit. To give a
population of light particles we must instead choose a trajectory (such as (e)) 
that curves downwards, driving the heavy particles to the trap center.
The congregation of the heavy particles at the trap center could
be monitored using density contrast imaging so would give a clear signature 
of ferromagnetic ordering.

\subsubsection*{Cloud size}

We now use the intuition developed from studying the density profiles
to explore the variation of the cloud size. We first focus on the
behavior of the atoms in a non-interacting and also a strongly
interacting cloud. Secondly, we will study two important features that
appear at intermediate interaction strengths: the emergence of a local
maximum in cloud size, and a gradient discontinuity in the cloud size.

Throughout our study of mass imbalance we have opted to keep the mass
of the light species constant and increase the mass of the heavy
species. As we see in \figref{Flo:traj}(a) at zero and weak
interactions the lighter species is the outermost in the trap whenever
the population is not strongly biased towards the heavy
particles. Therefore in \figref{Flo:array}(a and b), as we increase
mass imbalance the cloud size of the non-interacting gas is always the
same. However, in \figref{Flo:traj}(f), we see that if there is
sufficient population imbalance towards the heavier species, they can
instead be the outermost species. The crossover can be deduced from
the exact expression for the cloud size in the non-interacting limit
$3^{1/6}2^{5/12}\hbar^{1/2}\omega^{-1/4}\max_{\sigma}(N_{\sigma}^{1/6}/m_{\sigma}^{1/4})$.

We now turn to study the opposite limit of a strongly interacting
gas. As seen in \figref{Flo:traj}(e and j), all the atoms are in fully polarized
domains so the cloud size plateaus as a function of interaction
strength. The heavy particles are found in the trap center and their
degeneracy pressure supports an outer shell of the light
particles. Therefore, if the mass of the heavier particles is
increased their density must increase to retain the same degeneracy
pressure $P\propto n^{5/3}/m$, thereby shrinking the cloud. We also note that the heavy
particles have a higher density than the light, and so biasing the
population towards the heavier species decreases the overall size of
the cloud.

After summarizing the behavior of the cloud size at weak and strong
interactions we are well positioned to highlight two non-monotonic
features that arise at intermediate interaction strengths: firstly a
gradient discontinuity seen in \figref{Flo:array}(c and d), and
secondly a local maximum at intermediate interaction strength in
\figref{Flo:array}(a and b).

\subsubsection*{Cloud size gradient discontinuity}

The discontinuity in the gradient of the cloud size is visible in
\figref{Flo:array}(c and d) at $k_{\text{F}}a\approx1$ and
$k_{\text{F}}a\approx2$ respectively. For weakly-interacting clouds
with a sufficient population imbalance towards the heavier species,
\figref{Flo:traj}(f) shows that the heavier
species extends to greater radii than the lighter species. However, at
strong interactions in \figref{Flo:traj}(j) the lighter species exists
exclusively at large radii. We monitor the expulsion of the light
atoms through the series of density profiles and trajectories \figref{Flo:traj}(g and
h). As the outside of the light particle cloud moves past the outside
of the heavy particle cloud it raises the rate at which the cloud
size increases, thus introducing the gradient discontinuity. The trajectories
(g) and (h) in \figref{Flo:traj} flip from downwards to upwards curvature at this point. The kink
is guaranteed to emerge if at zero interaction, the heavier species
persists to the largest radius, that is
$\mu_{\uparrow}>\mu_{\downarrow}$. This is equivalent to the condition
that
$N_{\uparrow}/N_{\downarrow}>m_{\uparrow}^{3/2}/m_{\downarrow}^{3/2}$.

\subsubsection*{Cloud size local maximum }

When the trapped gas has a mass imbalance in \figref{Flo:array}(a and
b), the cloud size has a local maximum with rising interaction
strength. Having earlier understood that cloud size increases with
$k_{\text{F}}a$ at small interactions, to study the emergence of a
local maximum in the cloud size we focus on why the cloud size
decreases on increasing the interaction strength from the putative
local maximum. In this limit the gas is almost fully phase separated,
having a central region made up entirely of heavy particles, followed
by a density discontinuity to $n_{\downarrow\text{c}}$ at some radius
$R_{\text{c}}$, outside which light particles dominate but contain a
small number, $\delta N_{\uparrow}$, of the heavier particles.

As $k_{\text{F}}a$ is increased between \figref{Flo:traj}(d and e), these outer heavy particles
are forced into the trap center. This leads to an increase in local
light particle Fermi energy of $g\delta N_{\uparrow}$, so a number
$\sim\delta N_{\uparrow}g\nu_{\downarrow}$ of lighter particles move
to the position of the shell from larger radii, where
$\nu_{\downarrow}$ is the density of states at the shell. This
expulsion occurs when $g\nu_{\downarrow}=1$, which is consistent with
the ferromagnetic transition, and so we deduce that $\delta
N_{\uparrow}$ lighter particles move in from larger radii to fill the
void left by the transfer of $\delta N_{\uparrow}$ heavier particles
to the center of the trap. With space cleared for light particles, the
cloud size falls by $\delta N_{\uparrow}/4\pi
R^{2}_{\text{c}}n_{\downarrow\text{c}}$.

However, a second counteracting effect occurs. As $\delta
N_{\uparrow}$ heavy particles are absorbed by the central phase region
its size increases, inflating the cloud by $\delta N_{\uparrow}/4\pi
R^{2}_{\text{c}}n_{\uparrow\text{c}}=\delta N_{\uparrow}
(m_{\downarrow}/m_{\uparrow})^{3/5} /4\pi
R^{2}_{\text{c}}n_{\downarrow\text{c}}$, where we have invoked pressure 
conservation at the boundary so
$n_{\uparrow\text{c}}=n_{\downarrow\text{c}}(m_{\uparrow}/m_{\uparrow})^{3/5}$.
In the presence of a mass
imbalance this expansion is less than the fall in size of the outer
particles, and so overall the cloud shrinks by $\delta
N_{\uparrow}(1-(m_{\downarrow}/m_{\uparrow})^{3/5})/4\pi
R^{2}_{\text{c}}n_{\downarrow\text{c}}$, thus forming the cloud size
local maximum. Finally we notice that with
mass balance $m_{\uparrow}/m_{\downarrow}=1$ we predict that there is
no fall in the cloud size upon approaching full polarization, which is
consistent with our plots in \figref{Flo:array}(a to d).

\subsubsection*{Kinetic energy}

The variation of kinetic energy with interaction strength in
\figref{Flo:array}(e to l) shows strong trends with changing
mass imbalance. When a heavier species is introduced, the kinetic
energy of that species at zero interaction strength falls as
$E_{\text{K}\uparrow}=2^{-13/6}3^{4/3}\omega^{1/2}\hbar
N_{\uparrow}^{4/3}/m_{\uparrow}^{1/2}$. \figref{Flo:array} (i
to l) shows that, for any given population imbalance, a greater heavy
particle mass leads to a lower kinetic energy at large $k_{\text{F}}a$.
This occurs because for nonzero mass imbalance, the
central heavy particle domain of the gas has a kinetic energy per particle $\sim
n_{\uparrow}^{2/3}/m_{\uparrow}$. However, increasing the mass of the
heavier species increases the central density to maintain pressure
support. Pressure balance at the interface of heavy and light phase
regions demands
$(n_{\uparrow}/n_{\downarrow})^{2/3}=(m_{\uparrow}/m_{\downarrow})^{2/5}$,
which suggests that the kinetic energy per particle is $\sim
n_{\downarrow}^{2/3}m_{\downarrow}^{-2/5}m_{\uparrow}^{-3/5}$. Since
the density distribution $n_{\downarrow}$ of the light particles varies slowly with
$m_{\uparrow}$, the heavy kinetic energy per particle falls with mass
like $\propto m_{\uparrow}^{-3/5}$.

\subsubsection*{Loss rate}

In a study of three-body losses Ref.~\cite{Petrov03} in the presence
of mass imbalance it was found that the loss process is greatly
suppressed for large mass imbalance. We see in \figref{Flo:array}(m to
p) that clouds with the largest mass imbalance
($m_{\uparrow}/m_{\downarrow}=20/3$) have significantly reduced
three-body losses. Experiments on clouds with a lower loss rate will
have longer to reach equilibrium and so could potentially better
reflect theoretical predictions. Moreover, mass imbalance also drives
a double maximum in loss rate against $k_{\text{F}}a$, for example,
when $m_{\uparrow}/m_{\downarrow}=3/2$ in
\figref{Flo:array}(p). We now explore this feature
using our trap profiles in \figref{Flo:traj}. Increasing the
interaction strength from zero intuitively leads to an initial
increase in loss rate $\Upsilon n_{\uparrow}n_{\downarrow}a^{6}$. At
the interaction strength for \figref{Flo:traj}(g), the light
particles are expelled to larger radii where heavier particles are
less dense, so the three-body loss rate falls. However, following this
as the interaction strength is increased still further the loss rate
$\propto(k_{\text{F}}a)^{6}$ rises again until the interaction
strength is sufficient in \figref{Flo:traj}(j) to finally completely
expel the light particles out of the heavy particle region. At this
point the loss rate falls completely to zero. This system therefore
offers a fully polarized cloud, that is also stable to three-body
losses, which is not seen with mass balance since even at high
interaction strength the gas is always paramagnetic in the outer
regions thus giving a finite loss rate.

There is also a two-body loss process~\cite{Pekker10} that offers a competing many-body
instability to the Feshbach molecules seen on the BEC-BCS crossover. Though
the instability appears
to be important in the equal mass case, in the presence of population or mass imbalance
it is known that the superfluid gap is reduced~\cite{Baarsma10}. Therefore,
we expect that the two-body loss rate should also fall.

\section{Textured phases \& perpendicular magnetization}\label{sec:TextureandPerp}

We now study the stability of our uniform mean-field states to two
kinds of perturbation. Firstly, we address the possibility for
spontaneous in-plane polarization. Secondly, we study the opportunity
for a spin spiral state to emerge as a ground state instability of the
imbalanced Fermi seas. Thirdly, in the recent experimental study one
tactic to minimize three-body losses was to rapidly ramp the
interaction strength, so we search for the most unstable collective
modes following a quench. All three of these instabilities can be
studied through the magnetic susceptibility, which we first derive
below.

Starting with \eqnref{eq:phiaction}, we expand the magnetization
fields in small perturbations
$\delta\vecgrk{\phi}^{\omega,\mathbf{q}}$ around a stationary and
homogeneous saddle-point solution $\overline{\vecgrk{\phi}}$. From
\secref{sec:GrandCanonical} the mean-field $\overline{\vecgrk{\phi}}$
is aligned along the z-axis. There are no linear terms in
$\delta\vecgrk{\phi}^{\omega,\mathbf{q}}$, so to second order we get a
change in the action of
\begin{align}
 \delta S=&\,g\sum_{\omega,\mathbf{q}}\left|\delta\phi_{\perp}^{\omega,\mathbf{q}}\right|^{2}\left[1+\frac{g}{2}\left(\Pi_{\uparrow\downarrow}^{\omega,\mathbf{q}}+\Pi_{\downarrow\uparrow}^{\omega,\mathbf{q}}\right)\right]\nonumber\\
 +&\,g\sum_{\omega,\mathbf{q}}\left|\delta\phi_{\text{z}}^{\omega,\mathbf{q}}\right|^{2}\left[\frac{1-g^{2}\left(\Pi_{\uparrow\uparrow}^{\omega,\mathbf{q}}\Pi_{\downarrow\downarrow}^{\omega,\mathbf{q}}\right)}{1-\frac{g}{2}\left(\Pi_{\uparrow\uparrow}^{\omega,\mathbf{q}}+\Pi_{\downarrow\downarrow}^{\omega,\mathbf{q}}\right)}\right]\punc{,}
 \label{eq:dispersion}
\end{align}
where $\Pi_{\alpha\gamma}^{\omega,\vec{q}}=\frac{1}{\beta V}\sum_{\omega',\vec{q}'}[\cmplxi\omega'-\overline{\xi}_{\alpha}(\vec{q}')]^{-1}[\cmplxi\omega'+\cmplxi\omega-\overline{\xi}_{\gamma}(\vec{q}'+\vec{q})]^{-1}$,
and $\overline{\xi}_{\sigma}(\vec{p})=p^{2}/2m_{\sigma}-\mu_{\sigma}+gn_{-\sigma}$. To recover the Stoner
criterion we examine the $\omega=0$ and $\vec{q}=\vec{0}$ z-channel where $\Pi_{\sigma\sigma}^{0,\vec{0}}=-\nu_{\sigma}$, and the density of states $\nu_{\sigma}$ is evaluated at the $\sigma$ species Fermi surface. This then gives $\delta S=g|\delta\phi_{\text{z}}^{0,\mathbf{0}}|^{2}(1-g^{2}\nu_{\uparrow}\nu_{\downarrow})/[1+g(\nu_{\uparrow}+\nu_{\downarrow})/2]$, which has a ferromagnetic instability at $g\sqrt{\nu_{\uparrow}\nu_{\downarrow}}=1$.

\subsection{Perpendicular polarization}\label{sec:PerpPol}

An SU(2) spin symmetric system can become polarized in any
direction. However, if SU(2) symmetry is broken through mass and/or
chemical potential imbalance then numerics show that
$\phi_{\perp}=0$. Here we verify this result analytically by showing
that a system strongly polarized along the quantization axis is always
stable against the formation of perpendicular polarization $\phi_{\perp}$. 

Starting from \eqnref{eq:dispersion}, the system is stable
against in-plane polarization only if
$\eta\equiv1+g\Pi_{\uparrow\downarrow}^{0,\mathbf{0}}>0$. At zero
temperature, we find that
\begin{align}
 &\Pi_{\uparrow\downarrow}^{0,\mathbf{0}}=-\frac{m^{3/2}}{\sqrt{2}\pi^{2}\hbar^{3}}\frac{1-r^{2}}{r}\Biggl[\sqrt{\epsilon_{\text{F}\uparrow}(1+r)}-\sqrt{\epsilon_{\text{F}\downarrow}(1-r)}\neweqnline
 & +\Gamma \arctan\left(\frac{\sqrt{\epsilon_{\text{F}\downarrow}(1-r)}-\sqrt{\epsilon_{\text{F}\uparrow}(1+r)}}{\Gamma+\Gamma^{-1}\sqrt{\epsilon_{\text{F}\downarrow}\epsilon_{\text{F}\uparrow}(1-r^{2})}}\right)\Biggl]\punc{,}
  \label{eq:perpsusceptmain}
\end{align}
with $\Gamma=[\Delta \mu +g \phi_{\text{z}})(1-r^{2})/r]^{1/2}$, and
$\epsilon_{\text{F}\sigma}=\max\left(0,\mu_{\sigma}-gn_{-\sigma}\right)$. In the presence of population and mass balance we find that $\eta=0$ in the polarized regime ($g>1/\nu$). Therefore SU(2) symmetric systems are susceptible to transverse polarization.

We now show that perpendicular magnetization cannot spontaneously
develop when a mass or population imbalanced system is strongly
polarized along the z axis. An instability only emerges if $\eta$ turns
negative, so our strategy is to bound $\eta$ from below. Without loss
of generality we focus on the $\uparrow$ spin polarized system. $\eta$
decreases with $g$ right up to the fully polarized boundary so we
substitute in the value of $g$ at full polarization given by
\eqnref{eq:gfp}. This transforms $\eta$ into an increasing function of
$\Delta\mu/\mu$, so we use the smallest value of $\Delta\mu/\mu$ consistent with $\uparrow$ spin
polarization given by \eqnref{eq:implicit}. This allows us to bound
$\eta$ from below by
\begin{align}
 \eta\geq\frac{36r^{2}}{175}-\frac{8r^{3}}{2625}>0 \text{ for
 }0<|r|\leq1\punc{.}  \label{eq:perpineq}
\end{align}
The increasing powers of $r$ come from solving \eqnref{eq:implicit}
for $\Delta\mu/\mu$ using series. As $\eta>0$ the system is stable
against perpendicular polarization if the SU(2) symmetry is broken by
mass or population imbalance. Furthermore the perpendicular magnetization
fluctuations have a gapped spectrum, whereas in the mass balanced case they are soft~\cite{Conduit08}. 

\subsection{Textured phases at mean-field level}\label{sec:TexturedPhase}

To verify that the system is not unstable to the formation of a
textured phase we first study a spin spiral with polarization along
the quantization axis. We focus on long wavelength spirals so expand
in small $q\ll p_{\text{F}}$ to find that the longitudinal
susceptibility is
$\Pi_{\sigma\sigma}^{0,\vec{q}}=\frac{m_{\sigma}}{2\pi^{2}}(-p_{\text{F}\sigma}+q^{2}/12p_{\text{F}\sigma})$.
Substituting this into \eqnref{eq:dispersion} one finds the resulting
coefficient of $q^{2}$ is always positive, and so a textured phase only
serves to increase the coefficient of
$|\delta\phi_{\text{z}}^{0,\vec{q}}|^{2}$. Therefore a spin
spiral state is less energetically favorable than a
uniform ferromagnetic state. We have also verified that a similar
argument holds for spin spiral phases with in-plane polarization.

\subsection{Dynamical phase formation}\label{sec:DynamicalPhaseFormation}

\begin{figure}
 \includegraphics[width=0.9\linewidth]{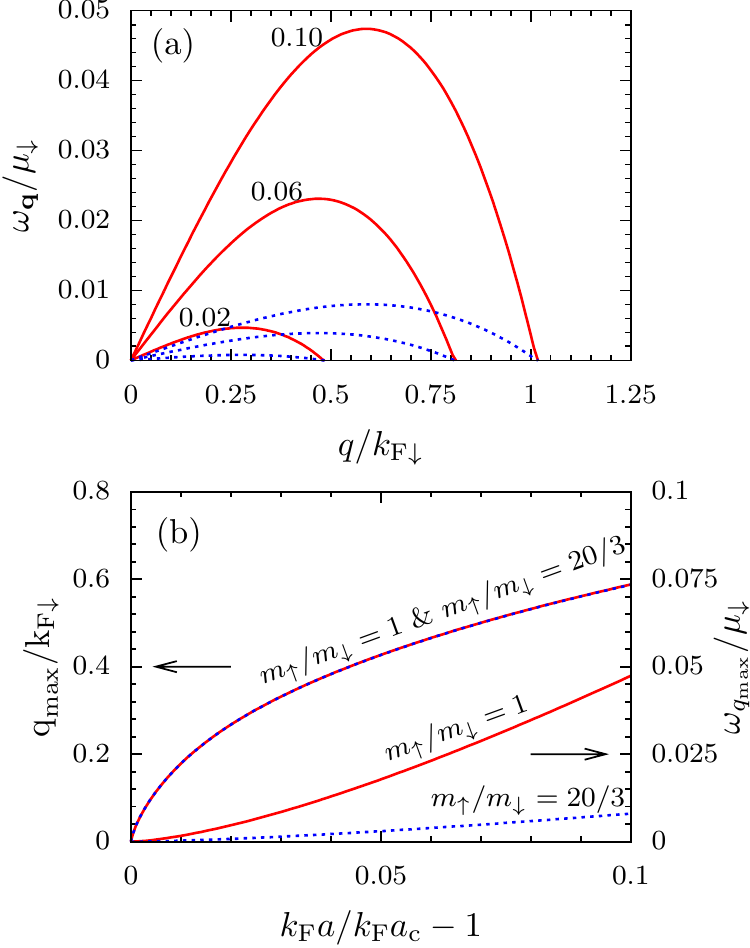}
 \caption{(Color online) (a) shows the growth rate $\omega_{\mathbf{q}}$ as a
 function of wave vector $q$ for collective modes at values of the
 dimensionless interaction strength
 $k_{\text{F}}a/k_{\text{F}}a_{\text{c}}-1=\{0.02,0.06,0.10\}$. 
 Here $k_{\text{F}}a_{\text{c}}$ is the critical interaction strength
 of the Stoner transition, given in the general mass and population
 imbalanced case by the boundary in \figref{Flo:2dzphase}. The
 solid red curves correspond to the mass balanced gas, and the dashed blue
 lines to a gas of $^{6}$Li and $^{40}$K, where the
 chemical potentials have been tuned to give population balance at
 $k_{\text{F}}a=0$. (b) Summarizes the behavior of mode wave vector
 (primary y-axis) and maximum growth rate (secondary y-axis) in (a).
 The axes are normalized
 by the Fermi wave vector and chemical potential of the $^{6}$Li atoms.}
 \label{Flo:disp}
\end{figure}

One tactic to reduce three-body losses in the ferromagnetic gas is to
study the dynamics immediately following an interaction strength
quench. The mass balanced ultracold atomic gas is predicted to form
unstable collective modes~\cite{Babadi09}, and here we explore the
mass imbalanced case. To obtain the wave vector $\vec{q}$ and growth
rate $\omega_{\vec{q}}$ of the unstable modes, we search for the poles in the
magnetization propagator, which are the solutions to
\begin{align}
\left(\mathrm{z}\right):\;1-g^{2}\Pi_{\uparrow\uparrow}^{\omega,\mathbf{q}}\Pi_{\downarrow\downarrow}^{\omega,\mathbf{q}}& =0\nonumber\\
\left(\perp\right):\;1+\frac{g}{2}\left(\Pi_{\uparrow\downarrow}^{\omega,\mathbf{q}}+\Pi_{\downarrow\uparrow}^{\omega,\mathbf{q}}\right) & =0\punc{.}
\label{eq:dynamics}
\end{align}
The growth rate relation for modes along the quantization axis is
given by solving $\left(\text{z}\right)$, whereas the relation for
perpendicular modes is given by solving $\left(\perp\right)$. The
polarization $\Pi_{\alpha\beta}^{\omega,\vec{q}}$ is as defined
immediately below \eqnref{eq:dispersion}, except with $g=0$.

We calculate these susceptibilities computationally. In
\figref{Flo:disp}(a) we plot the frequency against wave vector
relations for the z direction, and compare those with the mass balanced
case. The curves in \figref{Flo:disp}(a) reveal the most unstable mode
with largest growth rate $\omega_{\vec{q}}$ has a well defined maximum
at wave vector $q=q_{\max}$. In the context of experiment, we expect the
domains to be roughly of size $\sim 1/q_{\max}$, and have growth rates
$\sim \omega_{q_{\max}}$.

We see in \figref{Flo:disp}(b) that, for a particular normalized value
of the interaction strength, using the $^{6}$Li-$^{40}$K mass
imbalanced system does not affect the size of z domains but does
reduce the rate of domain growth by a factor of $\sim 5$ relative to a
mass balanced gas. However, in \figref{Flo:array}(n) we see that
$^{6}$Li-$^{40}$K three-body losses were suppressed by a factor of
$\sim 20$ compared to a mass balanced gas. Therefore, for the same net loss
the domains in a mass imbalanced gas can undergo $\sim 4 = 20/5$ times the growth. For the
perpendicular direction, the introduction of mass imbalance decreases
domain size, while increasing formation time.

\section{Discussion}\label{sec:Conclusions}

An ultracold atomic gas of fermions with repulsive interactions offers
investigators a new flexible system in which to realize itinerant
ferromagnetism. Introducing a mass imbalance between the two spin
species drives new distinctive features in the experimental
observables of the cloud size, release energy, and loss rate that
should help better characterize the formation of a ferromagnetic
phase. Furthermore a mass imbalance can strongly suppress the
three-body loss rate that plagues the formation of the ferromagnetic
phase.

The presence of a mass imbalance also opens up new opportunities to
study collective phenomena beyond those that can be realized in the
standard Stoner model. Though we showed that a spin textured phase
analogous to the Fulde-Ferrell-Larkin-Ovchinnikov state in
superconductors is not formed at mean-field level, it has already been
established that fluctuations corrections drive its formation even in
a mass balanced gas. The presence of a mass imbalance will alter the
Fermi surface nesting and could pose an interesting direction for
future research.

\acknowledgments

We thank David Pekker, Vadim Puller, Ben Simons, the anonymous referee, and especially
Gyu-Boong Jo, Wolfgang Ketterle, and Joseph Thywissen for useful
discussions. CVK acknowledges the financial support of the EPSRC. GJC
received support from the Royal Commission for the Exhibition of 1851,
the Feinberg Graduate School, the Kreitman Foundation, and National
Science Foundation Grant No. NSF PHY05-51164.

\end{document}